\documentclass[conference]{IEEEtran}
\IEEEoverridecommandlockouts
\usepackage{cite}
\def\BibTeX{{\rm B\kern-.05em{\sc i\kern-.025em b}\kern-.08em
    T\kern-.1667em\lower.7ex\hbox{E}\kern-.125emX}}
    \usepackage{flushend}

\usepackage{amsmath}
\usepackage{amssymb}
\usepackage{tabularx}
\usepackage{lscape}
\usepackage{mathtools}
\usepackage{booktabs}
\usepackage{algpseudocode}
\usepackage{algorithm}
\DeclareMathOperator*{\minimize}{minimize}
\DeclareMathOperator*{\argmin}{arg\,min}

\newcommand{\cut}[1]{ }
\usepackage{enumitem}
\usepackage[normalem]{ulem}
\usepackage[dvipsnames]{xcolor}
\usepackage{xcolor}
\usepackage{color}
\usepackage{lipsum}
\usepackage{ulem}    
\usepackage[utf8]{inputenc}
\usepackage{authblk}

\begin{document}

\title{ROOM: Adversarial Machine Learning Attacks Under Real-Time Constraints}
% \title{Conference Paper Title*\\

\author[1]{Amira Guesmi %\thanks{A.A@university.edu}
}
\author[2]{Khaled~N.~Khasawneh %\thanks{B.B@university.edu}
}
\author[3]{Nael~Abu-Ghazaleh %\thanks{C.C@university.edu}
}
\author[1]{Ihsen Alouani %\thanks{D.D@university.edu}
}

\affil[1]{IEMN-DOAE, UMR CNRS 8520 Polytechnic University Hauts-De-France, France}
\affil[2]{George Mason University, USA}
\affil[3]{University of California Riverside, USA}

\maketitle

\begin{abstract}
Advances in deep learning have enabled a wide range of promising applications.  However, these systems are vulnerable to {\em Adversarial Machine Learning (AML)} attacks; adversarially crafted perturbations to their inputs could cause them to misclassify.   
% adversarial noise undermines their trustworthiness. 
%Adversarial attacks are generated by a malicious actor by crafting an additive noise to the input which can force a classifier's output to a wrong label.  
Several state-of-the-art adversarial attacks have demonstrated that they can reliably fool classifiers making these attacks a significant threat.    %Moreover, as defenses are proposed, often new algorithms emerge that bypass these defenses.%jeopardize intelligent systems' integrity. 
Adversarial attack generation algorithms focus primarily on creating successful examples while controlling the noise magnitude and distribution to make detection more difficult.  The underlying assumption of these attacks is that the adversarial noise is generated offline, making their execution time a secondary consideration. However, recently, just-in-time adversarial attacks where an attacker opportunistically generates adversarial examples on the fly have been shown to be possible.  As an example scenario, an attacker observes the start of a command provided to Alexa and on the fly predicts the remainder of the command and generates an adversarial attack to cause Alexa to execute a different command.  %bringing adversarial attacks to the Edge, and dealing with real-time settings opens a new perspective in crafting, evaluating and defending against these attacks.  
This paper introduces a new problem: how do we generate adversarial noise under real-time constraints to support such real-time adversarial attacks?  Understanding this problem improves our understanding of the threat these attacks pose to real-time systems and provides security evaluation benchmarks for future defenses. Therefore, we first conduct a run-time analysis of adversarial generation algorithms. Universal attacks produce a general attack offline, with no online overhead, and can be applied to any input; however, their success rate is limited because of their generality.  In contrast, online algorithms, which work on a specific input, are computationally expensive, making them inappropriate for operation under time constraints. Thus, we propose ROOM, a novel Real-time Online-Offline attack construction Model where an offline component serves to warm up the online algorithm, making it possible to generate highly successful attacks under time constraints. Our results show that ROOM can achieve high attack success rates under real-time constraints with up to $90$ times faster adversarial attack generation than the state-of-the-art methods. For example, ROOM achieves $100\%$ adversarial attack success rate on MNIST data with a throughput of up to $1250$ frame per second (FPS), more than $60\%$ success rate with $200$ FPS on CIFAR-10 and $60\%$ with $16$ FPS on ImageNet. %We notice also that ROOM is $6.5\times$ more effective than YOPO for a throughput of 25 FPS on MNIST and $2\times$ more effective than YOPO for 12 FPS throughput on CIFAR-10. \ihsen{other examples from other dartasets-- exemples with yopo also?}

%similar to the state of the art attacks while significantly speeding up the run-time (online) attack generation; up to $5$ times faster than C\&W and up to  $12 $ times faster than PGD.   %Specifically, we investigate state-of-the-art attacks under online and offline time budgets along with noise budget constraints. \nael{Too short on our contribution -- perhaps we can talk about our contribution as a new algorithm that tries to warm up the online component.}
\end{abstract}

\begin{IEEEkeywords}
Adversarial machine learning, adversarial attack, real-time, Security
\end{IEEEkeywords}

\section{Introduction}\label{sec:intro}

The emergence of deep learning is causing disruptive transformations in a wide range of sectors such as computer vision \cite{simonyan2014deep,redmon2016yolo9000}, natural language processing (NLP) \cite{deng2018deep}, robotics \cite{pierson2017deep}, autonomous driving \cite{al2017deep}, and healthcare \cite{miotto2018deep}. %In fact, recent advances in machine-learning, especially in deep neural networks (DNN), have significantly accelerated the development and deployment of data-driven designs with increasingly high efficiency. 
While these technologies are already in use in products and systems enabled by the availability of increasing amounts of data, they suffer from vulnerabilities to adversarial attacks that threaten their integrity and their trustworthiness. In particular, Adversarial Machine Learning (AML) attacks modify an input to a machine learning classifier with carefully crafted perturbations chosen by a malicious actor to force the classifier to a wrong output. If attackers are able to manipulate the decisions of a machine learning classifier to their advantage, they can jeopardize the security and integrity of the system, and even threaten the safety of people that it interacts with. For example, adding adversarial noise to a stop sign that leads an autonomous vehicle to wrongly classify it as a speed limit sign~\cite{b0,papernot2016transferability,fgsm} potentially leading to crashes and loss of life. In addition, adversarial examples have been shown effective in real-world conditions~\cite{physical}: that when printed out, an adversarially crafted image can remain adversarial to classifiers even under different lighting conditions and orientations. Therefore, understanding and mitigating these attacks is essential to developing safe and trustworthy intelligent systems.
%a thorough analysis and a deep understanding of the adversarial examples threat is of paramount importance to develop safe and trustworthy intelligent systems. 

Adversarial attacks have received considerable attention recently in the research community-directed.  Several adversarial generation algorithms have been developed~\cite{ C&W, fgsm, carlini_gift, pgd, CarliniW17}, often to bypass proposed defenses~\cite{distillation_SP}.  The threat model assumed by these systems is one where the attacker develops the adversarial example without any time constraint; thus, the proposed algorithms focus on maximizing the attack's success rate while minimizing the noise budget available to the attacker.  Minimizing the noise budget makes the perturbations injected by the attacker less detectable, both with respect to human perception as well as automated detection.    %In fact, the first conceptual success constraint of an adversarial noise is its imperceptibility, especially in image recognition applications. Since the perceptibility is a complex and subjective concept, the state-of-the-art adversarial attacks use the noise magnitude to craft and assess the efficiency of adversarial examples. 
Specifically, $L_p$-norm metrics have been widely utilised to measure noise magnitude, namely $L_0$, $L_2$, and $L_\infty$ \cite{C&W} (further details in Section \ref{sec:bckg}). 

 Machine learning classifiers are commonly used in intelligent systems where they receive inputs and react to them in real-time.  For example, several applications such as voice assistants (Apple Siri and Amazon Alexa), intelligent surveillance cameras \cite{11} and intelligent transportation systems \cite{16} operate on data in real-time, as they interact with the real-world.  In the context of such systems, pre-generated adversarial attacks are limited to be pre-deployed to try to interfere with the system opportunistically.   Often, these attacks also are limited to \textbf{universal} attacks that are designed to generalize across inputs and are therefore less effective than custom attacks~\cite{5,6,7}.  
 
  Recently, several just-in-time adversarial attacks have been proposed~\cite{RT_rnn, Gong2019RT}.  In these scenarios, an attacker predicts the arrival of an input (perhaps based on partial observation of a time-series or streaming input) and desires to generate an adversarial attack in real-time to inject the perturbations as the input arrives.  These attacks are possible on systems operating on streaming data or data that otherwise arrive progressively over time.  This could be streaming sensor data, audio or video data, and other time-series data (e.g., stock market data).  In such a setting, the attacker can predict a future input to the system based on the observed data so far and opportunistically generate and inject the adversarial perturbations. However, because existing input-specific attack generation algorithms are computationally expensive, they are not suitable for generating adversarial examples just-in-time when the data arrives rapidly (e.g., video and audio data).  % In such a situation, the attacker may be able to use an off-line generated universal attack but such attacks have limited effectiveness. However, they are unable to generate a custom attack because adversarial attack generation algorithms are computationally expensive. % However, recent work has shown that adversarial attacks can be deployed real-world scenarios. Since adding holistic noise to an image in real world settings is impractical, a different noise generation approach had to be implemented. In fact, designing physical noise in computer vision applications requires "adversarial patches" that are easily attached to physical objects such as clothing or road signs. Therefore, physical-world adversarial attacks have to be designed under the constraint of printable and localized adversarial noise \cite{5,6,7}. Notice that such adversarial examples are \textbf{universal}, and \textit{generated totally offline}. 
In this paper, we formulate a new problem of generating input-specific (or custom/non-universal) attacks under time constraints.   If such attacks are possible under tight time budgets, they enable highly dangerous just-in-time adversarial attacks, substantially expanding the threat of adversarial attacks to intelligent systems and setting new goals for effective defenses.  Existing adversarial attacks fall into two categories: 
\begin{itemize}
    \item Universal attacks: They are generated based on an optimization problem that is solved iteratively over a whole dataset instead of targeting a single input sample such as in the previous setting. These attacks generate a universal noise that does not need online exploration but assumes total access to the dataset of the victim system, and more importantly, an infinite offline exploration time; and
    \item Input-specific attacks:  These state-of-the-art attacks design an input-specific perturbation for a given input under the constraint of a noise magnitude budget. They typically assume no constraints in the attack generation time \cite{CarliniW17} \cite{pgd}.  The perturbation is added to the original sample, and the resulting adversarial example is fed to the victim DNN. The most efficient state-of-the-art attacks are iterative \cite{CarliniW17} \cite{pgd} and require a considerable amount of time to converge. Therefore, this setting is not practical for our scenarios where the attack must be generated under real-time constraints.
    %real-life situations where target inputs can stream online.
    
    \end{itemize}
    
    \cut{Particularly, when the target system requires streaming input, i.e., the input is continuously processed as it arrives at real-time. The following cases are illustrations of such application domains:
\begin{itemize}
    \item \textbf{Streaming environment perception.} Several application domains such as video surveillance, biometric authentication, object detection and recognition in autonomous vehicles, etc. operate necessarily online by processing the coming data on-the-fly. The threat model in which an attacker is able to inject noise within the whole input frame requires direct access to information between the sensor and the processing elements, which is unpractical. 
    
    \item \textbf{Financial Trading Systems.} Financial institutions make trading decisions using automatic machine learning algorithms based on a sequence of observations of the market behavior. Potential attackers may target the trading machine learning systems  by adversarially manipulating the market conditions. However, the attack can only add perturbations to future (yet to be observed) market parameters, e.g., using market perturbations, based on the current real-time observations. Therefore, an online infinite time is impractical in these settings.
    
  \item \textbf{Real-time Speech/Speaker Recognition Systems.} Machine learning based real-time speaker recognition, speech recognition and automatic translation systems have been adopted in a wide range of applications, including security-sensitive domains. A malicious actor may try to fool these systems by generating a carefully designed noise that is indistinguishable by the human ear. Since the adversarial noise is crafted to a given sample, assuming the adversarial noise generation online on-the-fly is impractical.
\end{itemize}

\noindent
\textbf{(ii) Universal patch generation attacks }  are dedicated to real-time settings. They are generated based on an optimization problem that is solved iteratively over a whole dataset instead of targeting a single input sample such as in the previous setting. These attacks generate a universal noise that does not need online exploration, but assume a total access to the dataset of the victim system, and more importantly an infinite offline exploration time. 
}

We observe that these two approaches represent two ends of the spectrum with respect to adversarial attack algorithms: \textit{(a)} the universal attacks are fully offline and not customizable to the input, and \textit{(b)} the input-specific algorithms are fully online, working with the input but not benefiting from any offline optimization opportunities.  We first analyze these algorithms from the perspective of not only the traditional metrics of attack success and noise budget but also from the perspective of run-time overhead, showing that they cannot achieve success under limited time budgets.  We then propose a new attack model, the Real-time Offline-Online Model (ROOM), that unifies the two adversarial attack classes and results in a fast (real-time) generation of input-specific attacks.  Specifically, ROOM uses an offline generation step to generate patches that serve as a warm-up of the online component. On the other hand, the online component specializes the offline generated patch to the current input.  By starting from this offline state, the online component can rapidly converge towards a successful attack, providing an input-specific attack within a limited time budget, i.e., faster attack generation than state-of-the-art input-specific attack algorithms.  Specifically, our results show that ROOM substantially outperforms state-of-the-art adversarial attack algorithms for the same online time budget. For the same accuracy, it can improve the convergence time of the conventional Carlini and Wagner (C\&W) algorithm and the Projected Gradient Descent (PGD) by up to $5$ times and $12 $, respectively.  We believe that ROOM is an important step towards producing adversarial attacks that can be deployed just-in-time. Furthermore, additional optimization opportunities are orthogonal to ROOM, including the use of hardware acceleration.%   In summary, the existing approaches of adversarial noise generation are corner cases from a time perspective: either an infinite offline time budget, or an infinite online time budget. In this paper, we propose an empirical analysis aiming to establish a rather continuous perspective in the adversarial noise generation process. Our analysis considers the problem of adversarial noise from a totally novel perspective by including a time budget concept in addition to the noise budget.\\ %The contributions 

\noindent
\textbf{Contributions.} In summary, the contributions of this work are: 
\begin{itemize}
    \item We introduce a new problem of generating efficient adversarial attacks under time constraints, i.e., a limited time budget.
    \item We contribute a time-aware characterization of adversarial attacks considering time in addition to traditional constraints of attack success and noise budget.
    %as a constraint in the generation process of adversarial noise. 
    %\item Weaddress the possibility of attack generation under time constraints beyond state-of-the-art costly universal attacks. 
    \item Based on the time analysis, we propose ROOM, a new real-time offline-online attack model that enhances adversarial attacks efficiency under real-time constraints. We show that optimizations of adversarial attacks under time constraints are possible, and we show that balancing offline and online generation processes can enhance the efficiency of adversarial attacks.  The proposed model unifies and generalizes existing algorithms that are either fully offline, or fully online.
    \item ROOM achieves real-time adversarial noise generation with a throughput of $1250$ FPS, $200$ FPS and $16$ FPS for MNIST, CIFAR10 and ImageNet, respectively. %compared to less than 28 FPS, 1.4 FPS, and 0.625 FPS when using conventional PGD attack to attain comparable attack success rates.

    %\item We achieve about $90$ and $12$ times speed up \ihsen{add real time results in FPS} compared to the original PGD attack with comparable attack success rate results on MNIST and CIFAR10, respectively. 
    %Combining ROOM with YOPO \cite{yopo} (YOPO-ROOM), we achieve higher attack success rate when tested against YOPO under the same time constraint.
    %\item For reproducible research, we release the complete source code of our methodology. \footnote{Omitted for blind review} %https://github.com/AG-X09/Offline-Online-Model-OOM-}%Ommitted for blind review }
  %  \item We introduce a new metric for robustness ... and a new metric for adversarial  attack efficiency under noise and time constraints 
\end{itemize}

\section{Background}\label{sec:bckg}
In this section, we present a brief background on the threat model and the generation process of adversarial examples, with an overview of the most widely used attacks in the literature.

\subsection{Threat Model}

\subsubsection{Attacker Knowledge}

We presume a white-box setting where the attacker is well aware of the parameters of the victim classifier and has direct access to the model gradient. This information is used by the attacker to construct adversarial examples.

\subsubsection{Adversarial Goal}
The objective of the attacker is to compromise the integrity of the victim model. The attack success Rate is defined as  (1 - Classification Accuracy) and represents the proportion of total perturbed images in a dataset for which the adversarial noise forces the model to output a wrong label. A lower classification accuracy corresponds to higher attack success rate. 
Adversarial goals can be divided into two categories:

\noindent\textbf{Untargeted adversarial attack}: The goal of a non-targeted attack is to slightly modify the source image so that it is classified incorrectly by the target model, without special preference towards any particular output.
\begin{equation}
    f(x) \neq f(x^*)
\end{equation}

\noindent\textbf{Targeted adversarial attack}: The goal of a targeted attack is to slightly modify the source image so that it is classified incorrectly into a specified target class $t$ by the target model.
\begin{equation}
    f(x^*) = t
\end{equation}

In this work, we consider the targeted adversarial attack setting. 

\subsubsection{Noise Budget}

%{\bf Distance Metrics}
The adversarial examples should be imperceptible by humans, and hence are constrained in amplitude. The noise budget is generally expressed in terms $L_p$-norm distance, mainly $L_0$, $L_2$, and $L_ \infty$: 

\begin{equation}
    \left\|x\right\|_p = \left( \sum^{n}_{i = 1} \left |x_i \right | ^{p} \right)^{\frac{1}{p}}
\end{equation}

% These metrics focus on different aspects of visual significance: 
% \begin{itemize}
%     \item \textbf{$L_0$} counts the number of pixels with different values at corresponding positions in the two images. 
%     \item \textbf{$L_2$} measures the Euclidean distance between the two images $x$ and $x^*$.
%     \item  \textbf{$L_ \infty$} measures the maximum difference for all pixels at corresponding positions in the two images.
% \end{itemize}

\subsection{Generating Adversarial Examples}
\noindent\textbf{Problem definition}
An adversary, using information learnt about the structure of the classifier, tries to craft perturbations added to the input to cause incorrect classification. For illustration purposes, consider a CNN used for image classification. Given an original input image $x$ and a target classification model $ f(.) $, the problem of generating an adversarial example $x^*$ can be formulated as a constrained optimization \cite{pbform}:

\begin{equation}
\label{eq:adv}
     \begin{array}{rlclcl}
        x^* = \displaystyle \argmin_{x^* }  \mathcal{D}(x,x^* ),\\  
         s.t.  ~ f(x^* ) = l^* ,  ~ l \neq l^* 
\end{array}
\end{equation}

Where $\mathcal{D}$ is a distance metric used to quantify similarity between two images and the goal of the optimization is to minimize the added noise, typically to avoid detection of the adversarial perturbations. $l$ and $l^*$ are the two labels of $x$ and $x^*$, respectively:  $x^*$ is considered as an adversarial example if and only if the label of the two images are different ($ f(x) \neq  f(x^*) $) and the added noise is bounded ($\mathcal{D}(x,x^*) < \epsilon $ where $\epsilon \geqslant 0 $).

%\subsection{Adversarial Attacks}
To solve this optimization problem, several approaches have been proposed in the literature from which present the most widely used attacks:

\noindent \textbf{Fast Gradient Sign Method (FGSM).} 
FGSM \cite{fgsm} is a single-step, gradient-based, attack. An adversarial example is generated by performing a one step gradient update along the direction of the sign of gradient at each pixel as follows:

 \begin{equation}
     x^* = x + \epsilon sign (\nabla_{x}J_{\theta}(x,y))
 \end{equation}
Where $\nabla J()$ computes the gradient of the loss function $J$ and $\theta$ is the set of model parameters. The $sign()$ denotes the sign function and $\epsilon$ is the perturbation magnitude. 

\noindent \textbf{Projected gradient descent (PGD).} 
PGD \cite{pgd} is a stronger iterative variant of the FGSM where the adversarial example is generated as follows:
 \begin{equation}
x^{t+1} = \mathcal{P}_{\mathcal{S}_x}(x^t + \alpha \cdot sign (\nabla_{x}\mathcal{L}_{\theta}(x^t,y)) )
 \end{equation}
Where $\mathcal{P}_{\mathcal{S}_x}()$ is a projection operator projecting the input into the feasible region $\mathcal{S}_x$ and $\alpha$ is the added noise at each iteration.
The PGD attack tries to find the perturbation that maximizes the loss of a model on a particular input while keeping the size of the perturbation smaller than a specified amount.  

\noindent \textbf{Carlini \& Wagner (C\&W).} 
This attack \cite{C&W} is one of the state-of-the-art attacks. This latter has 3 forms based on different distortion measures ($l_0, l_2, l_{\infty}$). In this work we only consider the $l_2$ form as it has the best performance. It generates adversarial examples by solving the following optimization problem: \vspace{-5pt}

 \begin{equation}
     \begin{array}{rrclcl}
        \displaystyle \minimize_{\delta} & \multicolumn{3}{l}{\left\Vert \delta \right\Vert_2 + c\cdot l(x+\delta)}\\
        \textrm{s.t.} & x + \delta \in [0,1]^n 
\end{array}
 \end{equation}
Where $\left \Vert \delta \right \Vert_2$ is the smallest perturbation measured by the $l_2$ norm that makes the model misclassify into another/target class. $l(\cdot)$ is the loss function reflecting the distance between the current situation and the objective of the attack defined as:

 \begin{equation}
l(x) = max(max_{i \neq t } \{Z(x)_i\}-Z(x)_t - \kappa )
 \end{equation}
 
Where $Z(x)$ is the output of the layer before the softmax called \textit{logits}. $t$ is the target label, and $\kappa$ is called the confidence, a hyper-parameter used to enhance the transferability of the output. 
An adversarial example is considered as successful if $max_{i \neq t}\{Z(x)_i\}-Z(x)_t \leq 0$. In the C\& W attack, the box constrained optimization problem $x + \delta \in [0,1]^n$ is turned to an unconstrained problem by replacing $\delta$ with $\frac{1}{2}(tanh(w)+1)-x$, where $w$ is a new optimizer ranging in $(-\infty , +\infty)$.

\section{Time-Aware Analysis of Adversarial Noise}

\subsection{Proposed Approach: Offline-Online Attack Model }

\begin{figure*}[!htp]
\centering
\includegraphics[width=2\columnwidth]{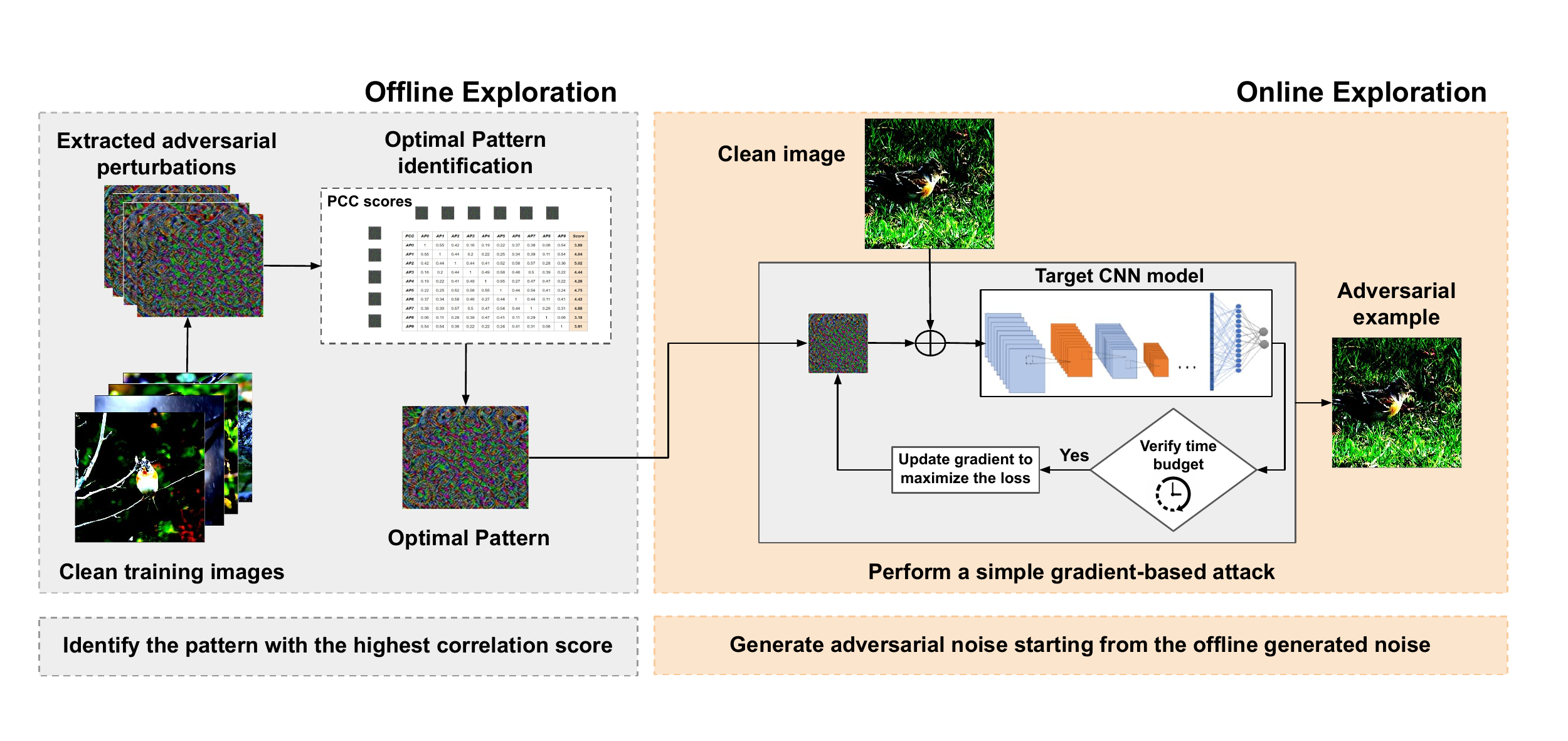}
\caption{Architecture of the proposed approach: Offline and Online exploration.}%\ihsen{notice that the time budget /real time is nonexistent in the figure} \amira{Done! please check} }
\label{process}
\end{figure*}

%\ihsen{Here I need to present a holistic view on the proposed analysis--- probably a figure? or a flowchart would help --- }

In contrast with the existing work on adversarial machine learning, this paper suggests including time as an analysis perspective of adversarial noise generation. More specifically, in addition to the noise budget used in the state-of-the-art, we consider \textbf{time budget} as an orthogonal constraint taken into account in the adversarial noise generation towards more practical threat models and defenses. 

%In this paper we introduce new metrics to the process of generating adversarial examples by including the \emph{time budget} as both the run-time and the design time.
To include time as a constraint in the adversarial noise generation process, we first distinguish the time budget from the two design spaces, i.e., online and offline, as follows:
\begin{itemize}
    \item \textbf{Online time budget:} which we note $T_{on}$, is defined as the time required for the online exploration, i.e., after the acquisition of the victim sample to target.
    \item \textbf{Offline time budget:} which we note $T_{off}$, is defined as the time required for the offline exploration to generate the adversarial perturbation. During this time, we assume that the attacker has no access to the victim data sample to target. 
\end{itemize}

Hence, we consider the adversarial noise generation as a continuous process combining offline and online processing. %, and we define the total time budget allocated to the generation of an adversarial example as:
% \begin{equation}\label{eq:time}
%   \delta T = T_{on} + T_{off}
% \end{equation}

We distinguish the two corner cases mentioned earlier in Section \ref{sec:intro}, where: (i) $(T_{off} = 0, T_{on}= \infty)$ corresponds to the conventional digital attack where all the computations are performed online without time limit,
and (ii) $(T_{off}= \infty , T_{on} = 0)$ corresponds to the universal adversarial perturbation referred to as Offline attack where all the computations are performed offline without time limit.
%Restricted RT: fix $\alpha$, vary $\beta$
%Restricted DT: fix $\beta$, vary $\alpha$

In this section, we propose real-time Offline-Online Model (ROOM), a new methodology for adversarial noise generation by combining both offline and online exploration under a time constraint. The proposed offline exploration is based on an analysis detailed later in Section \ref{sec:motiv}, where we explore the opportunities provided by patterns in adversarial examples that we exploit to warm-up the online exploration instead of starting from random initialization. Figure \ref{process} shows an overview on the proposed methodology and the two phases (offline and online) are detailed in Sections \ref{sec:off} and \ref{sec:on}. 

\subsection{Offline Exploration}\label{sec:off}
The main objective of the offline exploration is to identify the most efficient adversarial noise pattern that corresponds to a static adversarial component on which we can later (during the online exploration) build upon to quickly converge to an adversarial example. Algorithm \ref{offline_exp_algo} gives a detailed description of the offline exploration mechanism. To efficiently identify a potential noise pattern, the exploration process implements the following steps: First, we select a set of images for which we generate the corresponding adversarial examples (AEs) and collect adversarial perturbations $P_i$ (Line $4$ -- $7$ in Algorithm \ref{offline_exp_algo}). Next, we calculate the correlation between the resulting noise distributions (Line 9 -- 13). 
Subsequently, the aim is to identify the perturbation that has the highest correlation with the other samples, which correspond to the highest \emph{intra-class similarity} (detailed analysis in Section \ref{sec:motiv}). For this reason, we define the similarity score of a given noise as the sum of its PCCs with the remaining noise candidates:
\begin{align*}
    \mathcal{S}^{i}_{score} = \sum_{k \neq i} PCC_{i,k}
\end{align*}

Notice that the higher the similarity score, the closer the noise sample is to the potential static component (adversarial pattern) of the given intra-class setting. In fact, the noise candidate with the maximum $\mathcal{S}_{score}$ represents the highest static component that is redundant within most of the set's noise samples. Therefore, we finally identify the noise pattern as the noise candidate with the highest similarity score (Line 19--20).   %\ihsen{to be consistent with the motivation section, I need to keep calling the patch: pattern--}

%\ihsen{@Amira: in the algorithm, is the adversarial perturbation an input as stated?}

\begin{algorithm}
\caption{Offline Exploration (Optimal noise pattern identification) }
\label{offline_exp_algo}
\begin{algorithmic}[1]
\State \textbf{Input:} a classifier $f$, $x$ input image, $y$ true class, $y_{target}$ target class, noise budget $\varepsilon$, $m$ size of the used Set 
\State \textbf{Output:} $pattern_{adv}$
%\State Initialize~ $x_{adv} = \textit{perturbation}$
\State \textit{// Generate AEs and collect $P_i$}
\For {i = 0 \dots m-1}
\State $x^*_i = Attack(f, x_i, y_i, y_{target}, \varepsilon)$
\State $P_i = {x^*_i - x_i}$
\EndFor
\State \textit{// Calculate the correlation between patterns $P_i$ }
\For {s = 0 \dots m-1}
\For {k = 0 \dots m-1}
\State $PCC_{s,k} = corr(P_s, P_k)$
\EndFor
\EndFor
\State \textit{// Set a similarity score for each candidate pattern $P_i$ as the sum of PCCs }
\For {i = 0 \dots m-1}
\State $\mathcal{S}^{i}_{score} = \sum_{k \neq i}PCC_{i,k}$ %\ihsen{contradiction in the i index of the S -- @Amira please check this version--}
\EndFor
\State \textit{// Identify the $P_i$ with the highest correlation score}
\State $j = argmax(\mathcal{S}_{score}) $
\State $pattern_{adv} = P_j$
\end{algorithmic}
\end{algorithm}

%\begin{figure}
%\centering
%\includegraphics[width=\columnwidth]{figures/corr_score_mnist.PNG}
%\caption{Correlation scores (*)}
%\label{pcc_score_mnist}
%\end{figure}

%\begin{figure}
%\centering
%\includegraphics[width=0.9\columnwidth]{figures/cpp_score_mnist.PNG}
%\caption{Model accuracy when using patches with different correlation scores MNIST (*)}
%\label{pcc_score_mnist}
%\end{figure}

%\begin{figure}
%\centering
%\includegraphics[width=0.9\columnwidth]{figures/cpp_score_cifar.PNG}
%\caption{Model accuracy when using patches with different correlation scores CIFAR-10 (*)}
%\label{pcc_score_cifar}
%\end{figure}

\subsection{Online Exploration}\label{sec:on}

After the offline exploration, an adversarial noise pattern is identified as a potential static component that better characterizes the path "Source--Target" given a decision boundary defined by the trained victim classifier.  % The subsequent online step 
Our goal is to propose a novel approach for attacking trained models while considering a time budget. Hence, in the quest for a rapid adversarial example generation, the idea is to take advantage of the offline exploration to enhance the online generation efficiency. More specifically, we accelerate the conventional adversarial noise generation approaches with the offline-identified \emph{adversarial pattern}.
Essentially, the perturbation identified from the previous analysis is used as the initial starting point for a new adversarial attack targeting the same class. In fact, the noise pattern is identified such that it represents a static noise component that brings a given input from a source class $l$ closer to a target class $k$. Therefore, instead of starting the exploration of the adversarial noise from a zero or a random matrix, we start the online space exploration from an intermediate point that has a higher chance to be close to the decision boundary, and hence easier to flip the data sample classification to the target label. More importantly, the online exploration becomes faster in producing adversarial examples, allowing it to meet real-time constraints.   
% Based on the assumption that an adversarial perturbation for a previously attacked input will further guide the input to the direction of the targeted class better than starting from zero or random initial perturbation. 
Algorithm \ref{proposed_pgd} details an illustration of the online exploration where we use the proposed technique to build a Projected Gradient Descent (PGD) based attack where the adversarial example is initialized using the previously generated pattern.

%PGD is relatively fast (faster than C\&W) and the generated adversarial perturbations are small. C\&W is the most powerful attack but its computation is too heavy. Universal perturbation or Patches are the fastest but with high noise magnitudes..
%is a lightweight version of universal perturbation attacks with much lower design time where we identify a pattern between pairs of classes using targeted attacks. Use this pattern as a patch placed over the input image to generate adversarial examples.
%%Replace the attack with a simpler one while adaptively increase the perturbation rate.

\begin{algorithm}
\caption{Proposed PGD-ROOM attack}
\label{proposed_pgd}
\begin{algorithmic}[1]
\State \textbf{Input:} a classifier $f$ with loss $J$, noise budget  $\varepsilon$,step size $\alpha$, $x$ input image, $y_{target}$ targeted class, $pattern_{adv}$ adversarial pattern (identified offline), $m$ number of iterations, $T_{on}$ online time budget
\State \textbf{Output:} $x_{adv}$
\State $begin\_time = time.time()$
\State Initialize~ $x_{adv} = pattern_{adv}$
\For {i= 0...m-1}
    \If {$time.time()-begin\_time > T_{on}$}   %{prediction = $y_{target}$}
    \State break
    \EndIf
    
    \State prediction = $argmax(f(x_{adv})) $
    \State $x_{adv} = Clip \{x - \varepsilon sign (\nabla_{x_{adv}}J_{\theta}(f(x_{adv}),y_{target})) \}$
\EndFor
\end{algorithmic}
\end{algorithm}

%\begin{algorithm}
%\caption{Faster C\&W}
%\begin{algorithmic}[1]
%\State \textbf{Input:} a classifier $f$ with loss $J$, epsilon $\epsilon$, step size $\alpha$, $x$ input image, $y_{target}$ targeted class, $p$ adversarial perturbation, $T$ number of iterations 
%\State \textbf{Output:} $x_{adv}$

%\end{algorithmic}
%\end{algorithm}

%\subsection{New evaluation metrics}

\section{Experiments}

In this section, we evaluate the proposed methodology from different perspectives. Precisely, we first assess the offline exploration mechanism to verify the efficiency of the noise pattern identification, especially in terms of adversarial impact. Second, we investigate the behavior of ROOM under both time and noise constraints comparatively with conventional techniques. For thorough investigation, we compare ROOM with both zero and random initialization. Finally, we compare ROOM with a state-of-the-art adversarial training acceleration technique, i.e., YOPO, and show that combining ROOM and YOPO results in even further time efficiency, which offers an interesting property for low-cost adversarial training.

%we present our experimental setup and show the experimental results
%we first show adversarial examples generated by the proposed methods and analyze the properties of these examples from different perspectives:

\subsection{Setup}

Our experiments include implementations of a CNN architecture (Four convolutional layers and three fully connected layers) trained with MNIST \cite{mnist} for handwritten digit recognition. MNIST is composed of $60,000$ images, with $10$ classes and is composed of grey-scaled images of size $28\times28$ pixels.
We also use Wide ResNet-34 CNN trained on CIFAR-10 database \cite{CIFAR} for  object recognition in the evaluation. This database consists of 60,000 $32 \times 32$ RGB images in $10$ classes, with $6,000$ images per class. Finally, we consider VGG-19 CNN trained on ImageNet \cite{imagenet}, which contains over $14$ labeled million images of $224 \times 224$ pixels each.

To evaluate the attacks, we used two commonly used adversarial attack generation algorithms, namely:  %FGSM \cite{fgsm}, 
PGD \cite{pgd} and C\&W \cite{C&W}. 
%The FGSM is a commonly used baseline attack. 
The PGD attack is currently the strongest known attack for the $l_\infty$ metric. The C\&W attack is considered to be one of the strongest $l_2$ attacks. 
Our implementations are built using the open source machine learning framework PyTorch \cite{PyTorch}. We modified the FoolBox Library \cite{foolbox} to support our approach and to evaluate the attacks.
Experiments were taken on NVIDIA Tesla K80 GPU.

 %\ihsen{some details on the hardware we used maybe?}
%For the proposed attack, We first perform the offline exploration to generate the adversarial pattern, then use it as a starting point when performing the adversarial attack, i.e., the online exploration.

%---------------------------------------------------------
\subsection{Evaluation of the offline adversarial pattern exploration}
%---------------------------------------------------------
In this section, we evaluate the efficiency of the adversarial noise pattern extraction proposed in Section \ref{sec:off}.
For this reason, we use different patterns with different similarity score levels as the exploration initial starting point and monitor the classification accuracy of the model under attack. Specifically, we chose patterns with three similarity score levels: the minimum, the median, and the maximum among the exploration set. Recall that this latter corresponds to our choice of noise pattern when using the offline exploration. %\ihsen{did we mention this before?} \amira{the output of the offline exploration, last line in the offline exploration section}

\begin{figure}[!htp]
\centering
\includegraphics[width=0.8\columnwidth]{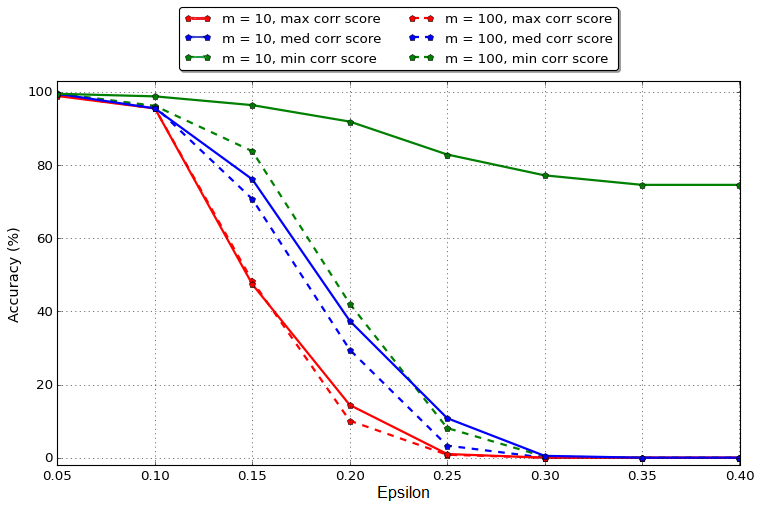}
\caption{Model classification accuracy under PGD attack when using patterns generated with max, median and min similarity scores for MNIST (m is the exploration Set size)}
\label{scores_mnist}
\end{figure}

\begin{figure}[!htp]
\centering
\includegraphics[width=0.8\columnwidth]{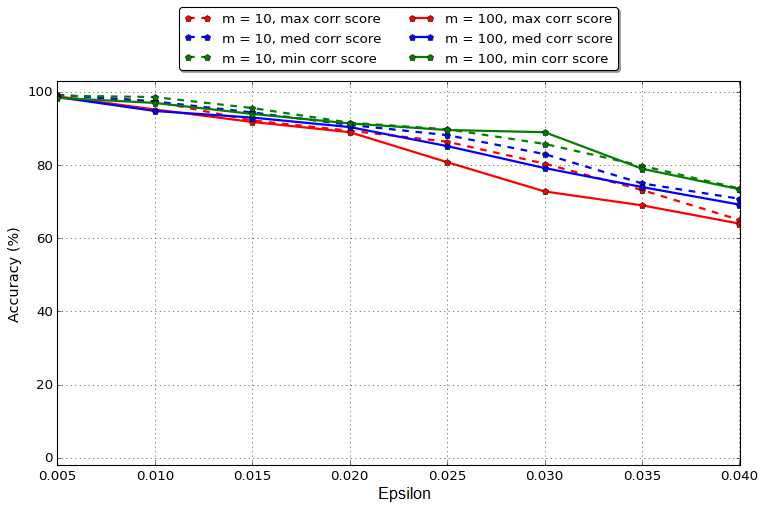}
\caption{Model classification accuracy under PGD attack when using patterns generated with max, median and min similarity scores for CIFAR-10 (m is the exploration Set size)}
\label{scores_cifar}
\end{figure}

Figures \ref{scores_mnist} and \ref{scores_cifar} show that using patterns with higher similarity scores leads to more powerful attacks. In fact, with a perturbation magnitude of $0.2$, we report a model classification accuracy degraded to less than $20\%$ when using the pattern with the highest score, while it remains up to $90\%$ when using the pattern with the lowest score. Changing the size of the exploration set ($m$) used to produce distinct adversarial patterns also reveals that the larger the employed set, the more likely it is to uncover a pattern with a higher similarity score and thereby a higher attack success rate. Notice that the $m$ samples of the considered sets are randomly chosen. 

%\ihsen{need to check m -- } done
%\ihsen{--I need to structure this--}
%\ihsen{more details needed }

%\ihsen{--- these are the offline experiments--- I need to  double check the sections--}

\subsection{Evaluation under noise and time constraint}
%---------------------------------------------------------
%\ihsen{work on the figures, make it more readable and time-focused\\}
In these experiments, we set a time budget for the online processing while varying the noise budget and comparing the performance of the state-of-the-art attacks to ROOM-enhanced version of the attacks.

\begin{figure*}[!htp]
\centering
\includegraphics[width=2\columnwidth]{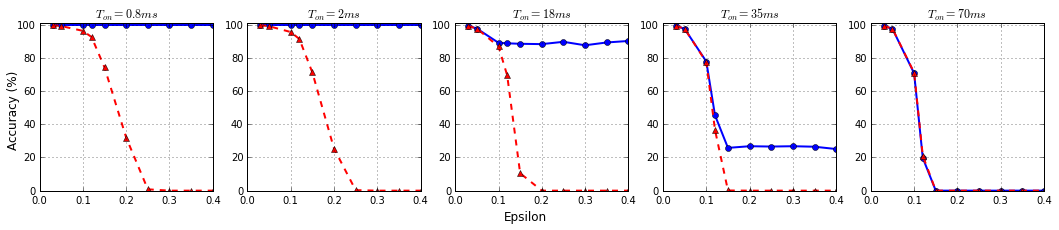}
\caption{Model classification accuracy under the PGD (blue) and the PGD-ROOM (red) attacks for different time budgets (MNIST).}
\label{pgd_mnist_acc}
\end{figure*}

%\begin{figure}[!htp]
%\centering
%\includegraphics[width=0.9\columnwidth]{figures/accuracy_runtime_rest_mnist.png}
%\caption{Model accuracy under the Online and the PGD-OOM attacks for different time budgets (MNIST).}
%\label{pgd_mnist_acc}
%\end{figure}

\noindent\textbf{Attacks on MNIST.} To assess the effectiveness of ROOM, we randomly choose four different pairs of (source, target) classes from MNIST test set and generate the optimal adversarial pattern for each to be used as initialization when generating the AE. 
In Figure \ref{pgd_mnist_acc}, we show the model accuracy under ROOM and conventional PGD attack comparatively for different noise budgets $\varepsilon$, and for different time budgets. %the average model classification accuracy under all PGD-based attack attempts is calculated. 

We can observe that PGD-ROOM efficiently generates adversarial examples at a maximum throughput of $1250$ FPS under a noise budget of $0.25$. More specifically, ROOM can \emph{totally} jeopardize the classification accuracy of the victim model processing data streaming in real-time with a speed of $1250$ FPS. However, at the same pace, the conventional online PGD attack was unable to have any impact on model accuracy even for a higher noise budget of $0.4$.

Moreover, for a throughput of $28$ FPS, the victim model \emph{misclassifies the totality of samples} under ROOM attack, while the baseline attack's maximum success rate is $75\%$ even for a higher noise budget.

%We can observe that for a $0.8$ ms 0.0008s time budget, PGD-ROOM delivers $0\%$ classification accuracy, which means a $100\%$ success rate of generating adversarial examples for a noise budget of $0.25$. However, no effect on the model accuracy was observed when using the conventional online PGD attack, even for a higher noise budget of $0.4$ under the same time constraint. 
%Moreover, PGD-ROOM is $90 \times$ faster than the conventional PGD, and for a time budget of $0.035$ seconds, the victim model misclassified all the samples, while the baseline attack's maximum success rate is $75\%$ even for a higher noise budget.

%}}
%The conventional attack takes $90 \times$ more time to reach the same performance.

%\begin{figure}[!htp]
%\centering
%\includegraphics[width=0.9\columnwidth]{figures/accuracy_runtime_rest_mnist_cw_v2.png}
%\caption{Model accuracy under the Conventional Online and the proposed C\&W-based attacks for different time budgets (MNIST).}
%\label{cw_mnist_acc}
%\end{figure}

\begin{figure*}[!htp]
\centering
\includegraphics[width=2\columnwidth]{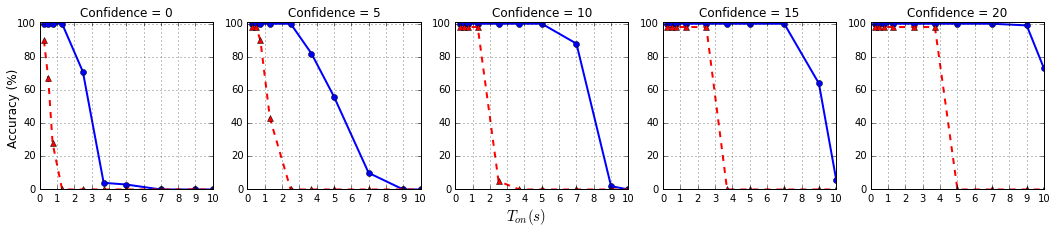}
\caption{Model classification accuracy under the C\&W (blue) and the C\&W-ROOM (red) attacks for different time budgets at various confidence levels (MNIST).}
\label{cw_mnist_acc}
\end{figure*}

In Figure \ref{cw_mnist_acc}, we evaluate the performance of the conventional C\&W comparatively to ROOM-C\&W. We measure the model accuracy under both attacks while varying the time budget for different confidence levels. Notice that \textbf{confidence} parameter for C\&W is the target confidence of the misclassification for an adversarial example. Thus, unlike the PGD attack that crafts adversarial examples within a given perturbation level, C\&W finds the smallest perturbation needed to cause misclassification with a given target confidence level.
Furthermore, higher confidence requires more time to be reached since it controls the gap between the generated AE and the decision boundary.
%For zero-confidence, under a $1.3$ seconds budget, C\&W-ROOM reaches $100\%$ success rate, while the conventional attack requires $5 \times$ more time to reach the same performance, which illustrates the effectiveness of our approach.

For zero-confidence, and with a throughput of $2$ FPS, C\&W-ROOM reaches $67\%$ success rate, while the model accuracy remains intact with the conventional attack. The conventional C\&W attack requires $5 \times$ more time to reach the same performance, which illustrates the effectiveness of our approach.

%\begin{figure}[!t]
%\centering
%\includegraphics[width=\columnwidth]{figures/accuracy_runtime_rest_cifar10.png}
%\caption{Model accuracy under the Conventional Online and PGD-OOM attacks for different time budgets (CIFAR-10).}
%\label{pgd_cifar_acc}
%\end{figure}

\begin{figure*}[!htp]
\centering
\includegraphics[width=1.6\columnwidth]{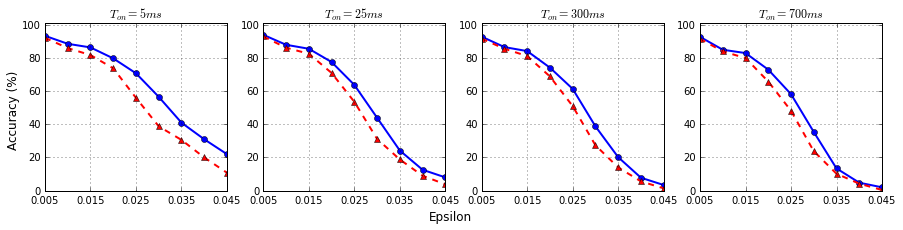}
\caption{Model classification accuracy under the PGD (blue) and PGD-ROOM (red) attacks for different time budgets (CIFAR-10).}
\label{pgd_cifar_acc}
\end{figure*}

\noindent\textbf{Attacks on CIFAR-10.} We generate adversarial examples for images from the CIFAR-10 dataset. Attacking deeper models and larger size and more complex images (RGB) requires a higher time budget. As shown in Figure \ref{pgd_cifar_acc}, ROOM outperforms the conventional PGD attack. 

\begin{figure*}[!htp]
\centering
\includegraphics[width=2\columnwidth]{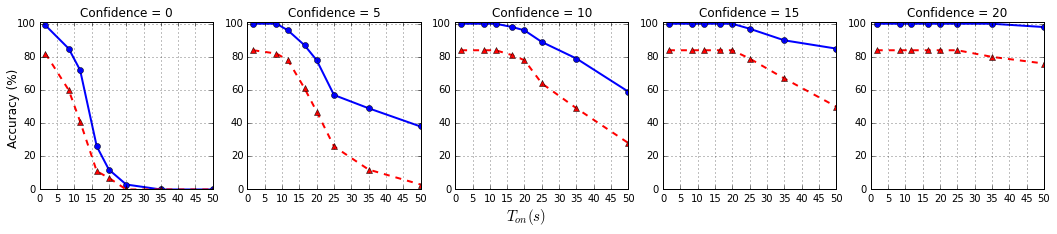}
\caption{Model classification accuracy under the C\&W (blue) and the C\&W-ROOM (red) attacks for different time budgets at various confidence levels (CIFAR-10).}
\label{cw_cifar_acc}
\end{figure*}

%Our proposed attack and the conventional attack delivers 
For the same target attack success rate of $60\%$ under a noise budget $\varepsilon = 0.03$, PGD-ROOM offers a throughput of $200$ FPS, while the maximum throughput of the conventional PGD is only $1.4$ FPS.
The same trend was observed for C\&W-based attacks: Figure \ref{cw_cifar_acc} shows clearly that ROOM outperforms the conventional attack. For instance, for a time budget of $10$ seconds, ROOM is $30\%$ more efficient than the conventional C\&W. For a confidence equal to $5$, the conventional attack takes $2 \times$ more time to reach the same success rate.

%For a time budget of $0.005$ seconds, PGD-ROOM is $10\%$ more effective than the conventional PGD attack. PGD-ROOM requires $12 \times $ less time than a conventional attack to reach the same performance. For example, for a noise budget of $0.035$ and a time budget of $0.03$ seconds, ROOM achieves an attack success rate of $85\%$.
%The same trend was observed for C\&W-based attacks; In Figure \ref{cw_cifar_acc}, we can clearly see that ROOM outperforms the conventional online attack.
%For instance, for a time budget of $10$ seconds, ROOM is $30\%$ more efficient than the conventional C\&W. For confidence equal to 5, the conventional attack takes $2 \times$ more time to reach the same success rate. %}}

%\begin{figure}[!htp]
%\centering
%\includegraphics[width=\columnwidth]{figures/accuracy_runtime_rest_cifar10_cw_v2.png}
%\caption{Model accuracy under the conventional C\&W and the C\&W-OOM attacks for different time budgets (CIFAR-10).}
%\label{cw_cifar_acc}
%\end{figure}

\begin{figure*}[!htp]
\centering
\includegraphics[width=1.6\columnwidth]{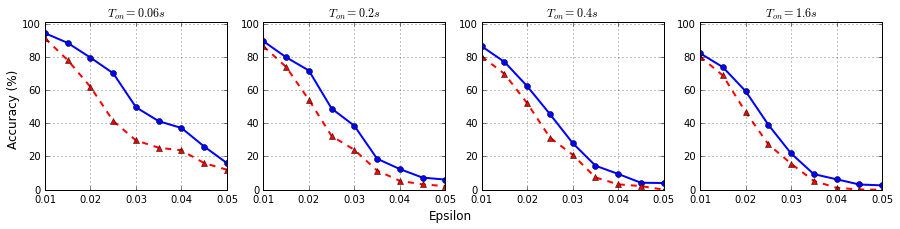}
\vspace{-8pt}
\caption{Model classification accuracy under the PGD (blue) and PGD-ROOM (red) attacks for different time budgets (ImageNet).}
\label{pgd_imagenet_acc}
\end{figure*}

\noindent\textbf{Attacks on ImageNet.}
In this section, we evaluate our approach on the ImageNet dataset. As shown in Figure \ref{pgd_imagenet_acc}, our proposed PGD-ROOM attack is found to be more effective. 
PGD-ROOM delivers the same attack success rate of $60\%$ for a throughput of $16$ FPS when the noise budget  $\varepsilon = 0.025$. However, the conventional attack achieves the same attack effectiveness for a throughput of $0.625$ FPS.

\begin{figure*}[!htp]
\centering
\includegraphics[width=1.3\columnwidth]{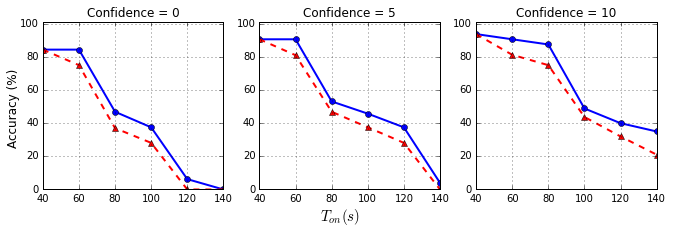}
\vspace{-8pt}
\caption{Model classification accuracy under the C\&W (blue) and C\&W-ROOM (red) attacks for different time budgets at various confidence levels (ImageNet).}

\label{c&w_imagenet_acc}
\end{figure*}

The same trend has been observed for C\&W-based attacks, as shown in Figure \ref{c&w_imagenet_acc}. C\&W-ROOM is $15\%$ more effective than the conventional C\&W attack for the same allocated time.

%In this section, we evaluate our approach on the ImageNet dataset. As shown in Figure \ref{pgd_imagenet_acc}, for the same allocated time budget, our proposed PGD-ROOM attack is found to be more effective. We can notice a $10\%$ difference in model accuracy for a time budget of $0.2$ seconds when the noise budget  $\varepsilon = 0.025$. ROOM is two times faster than the conventional attack.The same trend has been observed for C\&W-based attacks, as shown in Figure \ref{c&w_imagenet_acc}. C\&W-ROOM is more effective than the conventional C\&W attack and is $1.25 \times$ faster.

%\begin{figure}[!htp]
%\centering
%\includegraphics[width=\columnwidth]{figures/accuracy_runtime_rest_imagenet.png}
%\caption{Model accuracy under the Online and PGD-OOM attacks for different time budgets (ImageNet).}
%\label{pgd_imagenet_acc}
%\end{figure}

%\begin{figure}[!htp]
%\centering
%\includegraphics[width=\columnwidth]{figures/accuracy_runtime_rest_imagenet_cw_v2.png}
%\caption{Model accuracy under the Online and C\&W-OOM  attacks for different time budgets (ImageNet).}

%\label{c&w_imagenet_acc}
%\end{figure}

%---------------------------------------------------------
\subsection{Does ROOM have inherent impact on attack generation?} % vs zero and random noise initialization}
%---------------------------------------------------------

In this section, our objective is to investigate the impact of the offline exploration from a \emph{pure attack efficiency perspective} for a given noise budget. We want to answer whether ROOM's impact is inherent or can be achieved by any random initialization?

For a conclusive evaluation, we explore ROOM comparatively with both zero and random initialization of the adversarial noise. We use PGD with different numbers of steps.\\ %Notice that 
%We conduct experiments on MNIST and CIFAR-10 to demonstrate the effectiveness of OOM against other initialization methods such as zero and random noise initialization. 

\noindent\textbf{MNIST.} We set the size of perturbation as $\varepsilon = 0.3$ with a step size of $\frac{8}{255}$. We use PGD with different numbers of iterations and we report the classification accuracy in Table \ref{initial_MNIST}. We notice that ROOM outperforms all other approaches. For instance, with PGD-10 (PGD with $10$ iterations), ROOM is $3\times$ more powerful than random noise-based attack and more than $6\times$ more powerful than zero noise-based attack.

%MNIST, No Adversarial Training, step size = 0.01, epsilon = 0.3
%Clean data accuracy: 99.94\%
%network: 4 conv + 3 FC
%CIFAR-10, No Adversarial Training, step size = 0.002, epsilon = 0.03
%Clean data accuracy: 94.96\%
%network: Wide resnet34
\begin{table}[!htp]
\centering
  \caption{Model classification accuracy under PGD attack with different initialization methods on MNIST.}
  \label{initial_MNIST}
    %\resizebox{6cm}{!}{
  \begin{tabular}{cccccc}
    \toprule
       \textbf{Initialization}   & \textbf{PGD-40} & \textbf{PGD-10}  & \textbf{PGD-5} & \textbf{PGD-3} & \textbf{PGD-2} \\
    \midrule 
        Zero       &  1.6\%  &  84.9\% & 96.8\% & 98.5\% &99.5\%\\
        Random     &  0.2\% &  47\%  &  89.5\% &95.8\%&97.1\%\\  
        \textbf{ROOM}    &   \textbf{0\%}  & \textbf{13.7\%} & \textbf{47.7\%} & \textbf{69.4\%} & \textbf{76.3\%} \\  

  \bottomrule
\end{tabular} %}
\end{table}

\noindent\textbf{CIFAR-10.} We set the size of perturbation as $\varepsilon = 0.03$ with a step size of $\frac{2}{255}$. As shown in Table \ref{initial_CIFAR}, using ROOM made the attack more powerful, $1.5\times$ and $1.3\times$ more effective than zero and random noise-based attacks, respectively.

\begin{table}[!htp]
\centering
  \caption{Model classification accuracy under PGD attack with different initialization methods on CIFAR-10.}
  \label{initial_CIFAR}
    %\resizebox{6cm}{!}{
  \begin{tabular}{cccccc}
    \toprule
       \textbf{Initialization}   & \textbf{PGD-20} & \textbf{PGD-4}  & \textbf{PGD-3} & \textbf{PGD-2} & \textbf{PGD-1} \\
    \midrule 
        Zero       &  0\%  &  9.56\% & 19\% &40.87\% & 71.66\%\\
        Random     &  0\% &    8.8\% & 15.2\%&32.28\% & 62.1\%\\  
        \textbf{ROOM}    &   \textbf{0\%}  & \textbf{6.74\%} & \textbf{13.1\%} & \textbf{26.53\%} & \textbf{48.81\%} \\  

  \bottomrule
\end{tabular} %}
\end{table}

\noindent\textbf{ImageNet.} We use VGG-19 as the classifier for testing ImageNet. The noise magnitude is set to $\varepsilon = 0.03$ with step size $\frac{2}{255}$. As illustrated in Table \ref{initial_imagenet}, using ROOM with PGD attack made the attack more powerful, $1.6\times$ and $1.3\times$ more effective than zero and random noise-based attacks, respectively for PGD-3.

Those results confirm that our offline exploration significantly impacts an attack generation perspective even without considering time constraints. ROOM helps generate more effective adversarial attacks when compared to other initialization methods and for a limited number of attack iterations. %under tight time constraints.% attack Our technique achieves results superior to the conventional PGD attack when using both random and zero noise initialization on the MNIST, CIFAR-10 and ImageNet data sets. In fact, ROOM helps generate more effective adversarial attacks when compared to other initialization methods and for a limited number of attack iterations. %under tight time constraints.

\begin{table}[!htp]
\centering
  \caption{Model classification accuracy under PGD attack with different initialization methods on ImageNet.}
  \label{initial_imagenet}
    %\resizebox{6cm}{!}{
  \begin{tabular}{ccccc}
    \toprule
       \textbf{Initialization}   & \textbf{PGD-20} & \textbf{PGD-5}  & \textbf{PGD-3}  & \textbf{PGD-1} \\
    \midrule 
        Zero       &  0\%  &  15.62\% & 34.37\%  & 75\%\\
        Random     &  0\% &   12.5\% & 28.12\% &  71.87\%\\  
        \textbf{ROOM} & \textbf{0\%}  & \textbf{8.37\%} & \textbf{20.87\%} & \textbf{55.62\%} \\  

  \bottomrule
\end{tabular} %}
\end{table}

%---------------------------------------------------------
\subsection{ROOM vs YOPO}
%---------------------------------------------------------
In the quest of accelerating the adversarial training process, You Only Propagate Once (YOPO) \cite{yopo} has been recently proposed. YOPO is based on reducing the total number of full forward and backward propagation to only one for each group of adversary updates by restricting most of the forward and back propagation within the first layer of the network, taking advantage of the baseline training gradient backpropagation. 
%This approach was proposed as a solution to reduce the computational overhead of the generation of adversarial examples.
%based on the observation that the adversary update is only coupled with the parameters of the first layer of the network.
To evaluate our approach comparatively with YOPO, we run experiments where we set a time budget to $0.04$s and $0.08$s for adversarial example generation (corresponding respectively to $25$ and $12$ FPS) and we compare the effectiveness of each attack. Since ROOM is orthogonal to YOPO (YOPO focus on accelerating gradient-based noise generation, while ROOM focuses on noise patch initialization), we also explore the results of combining both techniques.

\noindent\textbf{MNIST.} As illustrated in Table \ref{yopo_mnist}, PGD-ROOM is more efficient in generating adversarial examples than YOPO initialized with zero and random noise for a throughput of 25 FPS. PGD-ROOM is nearly $6.5\times$ and $2.5\times$ more efficient than YOPO-Zero and YOPO-Random, respectively. Interestingly, we also noticed that combining YOPO with ROOM resulted in more gain in attack success rate.

\begin{table}[!htp]
\centering
  \caption{Model classification accuracy under ROOM vs YOPO on MNIST for a throughput of 25 FPS.}
  \label{yopo_mnist}
    %\resizebox{6cm}{!}{
  \begin{tabular}{ccc}
    \toprule
       \textbf{Attacks}   & \textbf{Eps = 0.2} & \textbf{Eps = 0.3}  \\
    \midrule 
        PGD-Zero       &  52.5\%  &  51.2\% \\
        PGD-Random     &  41.9\%  &  17.6\% \\  
        PGD-ROOM        &  23.2\%  &  6.8\%  \\  
        YOPO-Zero       &  44.8\%  &  44.6\% \\
        YOPO-Random     &  43\%    &  16\% \\  
        YOPO-ROOM        &  20.2\%  &  3.7\%  \\  
  \bottomrule
\end{tabular} %}
\end{table}

\noindent\textbf{CIFAR-10.} For a larger model, and a throughput of 12 FPS, we noticed the same trend; ROOM with PGD outperforms YOPO, and combining ROOM and YOPO yields to even better performance. For instance,  YOPO-ROOM is more than $2\times$ more powerful than YOPO-random, and PGD-ROOM is $1.4\times$ more successful than YOPO-random.

%\ihsen{Is it possible to have ImageNet ? }
\begin{table}[!htp]
\centering
  \caption{Model classification accuracy under ROOM vs YOPO on CIFAR-10 for a throughput of 12 FPS.}
  \label{yopo_cifar}
    %\resizebox{6cm}{!}{
  \begin{tabular}{ccc}
    \toprule
       \textbf{Attacks}   & \textbf{Eps = 0.02} & \textbf{Eps = 0.03}  \\
    \midrule 
        PGD-Zero       &  37\%  &  30\% \\
        PGD-Random     &  28\%  &  21\% \\  
        PGD-ROOM        &  15\%  &  12\%  \\  
        YOPO-Zero      &  23\%  &  19\% \\
        YOPO-Random    &  20\%  &  17\% \\  
        YOPO-ROOM       &  16\%  &  8\%  \\  
  \bottomrule
\end{tabular} %}
\end{table}

%\noindent\textbf{ImageNet.} For a more challenging dataset, the ImageNet, and under a time constraint set to $0$s, we noticed 

%\begin{table}[!htp]
%\centering
%  \caption{Classification accuracy ROOM vs YOPO on ImageNet for a time budget of 0.0s.}
%  \label{yopo_cifar}
    %\resizebox{6cm}{!}{
%  \begin{tabular}{ccc}
%    \toprule
%       \textbf{Attacks}   & \textbf{Eps = 0.02} & \textbf{Eps = 0.03}  \\
%    \midrule 
%        PGD-Zero       &  \%  &  \% \\
%        PGD-Random     &  \%  &  \% \\  
%        PGD-ROOM        &  \%  &  \%  \\  
%        YOPO-Zero      &  \%  &  \% \\
%        YOPO-Random    &  \%  &  \% \\  
%        YOPO-ROOM       &  \%  &  \%  \\  
%  \bottomrule
%\end{tabular} %}
%\end{table}

%\ihsen{Needs rewriting -- more material. We still have almost one page left.\\}

%In fact, we were able to achieve the same adversarial attack effectiveness but with a much fewer computational cost.

In conclusion, we notice that ROOM generates more effective adversarial attacks than YOPO under the same time constraint, which is a critical metric in scenarios like adversarial training when adversarial examples need to be generated for the whole training set.
Moreover, YOPO can be applied only to gradient-based attacks, while ROOM can be integrated into any attack generation method. %. For instance, ROOM also helped considerably accelerate the optimization-based C\&W attack.
Interestingly, since ROOM is orthogonal to YOPO, we noticed that combining both techniques results in an even more efficient adversarial example generation. We believe this is a promising property of ROOM to help provide more optimized AML protection.

\label{exp}

\section{How does the offline exploration accelerate adversarial noise generation?}\label{sec:motiv}
\begin{figure}[!t]
\centering
\includegraphics[width=\columnwidth]{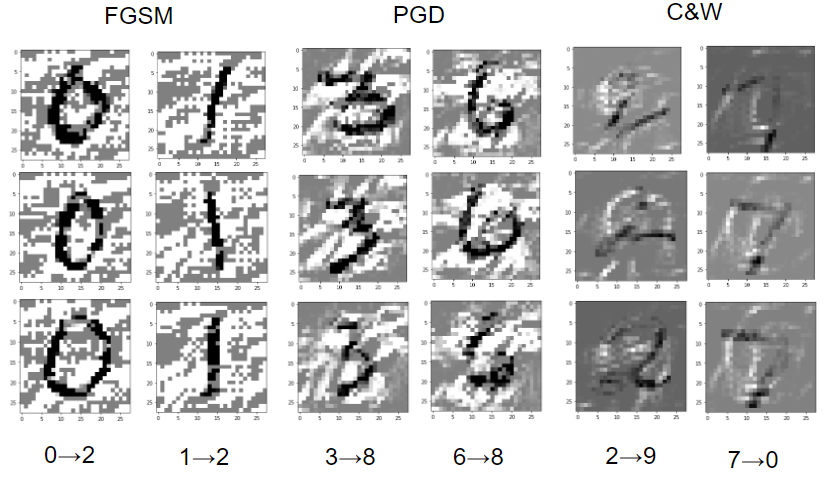}
\caption{Perturbation distributions samples under different targeted adversarial attacks on MNIST.}
\label{pattern}
\end{figure}

In this section, we investigate the spatial patterns of adversarial noise in image recognition applications. Conceptually, we explore if the noise resulting from pushing a given example towards the decision boundary contains a specific spatial pattern. This study provides the basis for the idea of splitting noise generation between offline and online phases towards a generic perspective of adversarial noise generation under noise and time constraints. 

\begin{figure}[!t]
\centering
\includegraphics[width=\columnwidth]{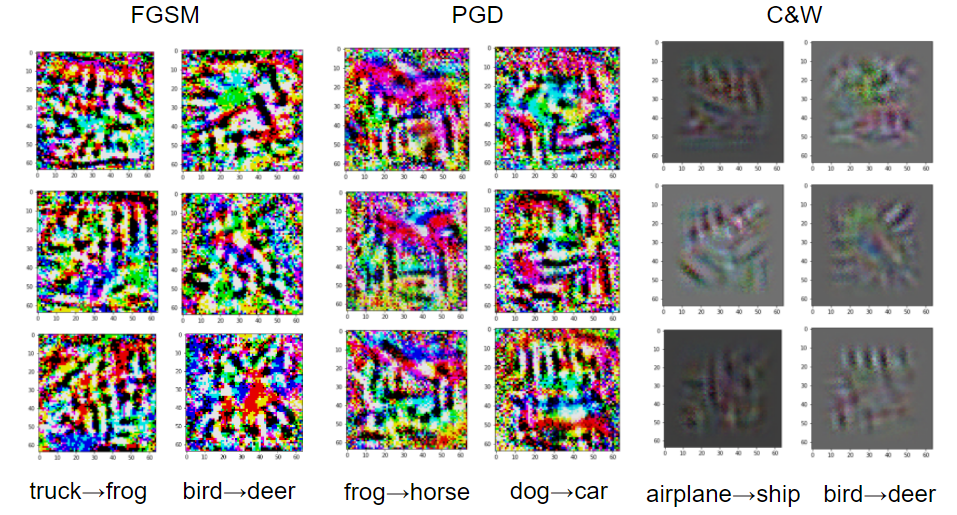}
\caption{Perturbation distributions samples under different targeted adversarial attacks on CIFAR-10.}
\label{pattern_cifar}
\end{figure}

First, we consider samples belonging to the same initial class on which we generate targeted adversarial noise towards a fixed target class. The objective is to explore the existence of potential spatial patterns in the noise distributions generated by adversarial attacks. An illustration of noise examples for different targeted attacks on MNIST and CIFAR-10 datasets are shown in Figures \ref{pattern} and \ref{pattern_cifar}, respectively. The illustrations show that a general pattern between pairs of classes could be extracted from the resulting adversarial noise. For MNIST dataset, since the data is relatively simple, we could see clear patterns for targeted adversarial attacks with plausible semantics; Attacking these pixels will alter the original handwriting shape towards the target class’s shape. For the CIFAR-10 dataset, since the images have three channels (RGB), the noise is not easily interpretable, yet, patterns can be identified for the same targeted attacks. 

To further assess the generalizability of this observation, we perform a statistical study on the similarity of generated noise within the same target and compare its distribution to adversarial noise for different targets.
To assess this noise matrix similarity, we use the Pearson correlation coefficient (PCC) \cite{pcc}, which is a widely adopted metric to measure the linear correlation between two variables.
% \ihsen{------ here comes the intuition, and prepare the proposed methodology-------}
% these patterns all perform well and could be utilized for computation optimization, which will be discussed in Section \ref{exp}. Same for images from ImageNet datasets. %\ihsen{}

%\subsection{Pearson Correlation Coefficient} %\ihsen{needs to go in the proposed approach, setup part--}

PCC coefficient is defined as follows:
\begin{equation}
    PCC_{X,Y} = \frac{cov(X,Y)}{\sigma_X \sigma_Y}
\end{equation}

where $cov$ indicates the covariance and $\sigma_X$ and $\sigma_Y$ are the
standard deviations of matrices $X$ and $Y$, respectively, and the $PCC$ values range from $-1$ to $1$. The absolute value indicates the extent to which the two variables are linearly correlated, with $1$ indicating perfect linear correlation, $0$ indicating zero linear correlation, and the sign indicates whether they are positively or negatively correlated.

The targeted adversarial noise is iteratively constructed to transform a sample from a source class (correct label) towards an adversarial sample $x^*_i$ with a target class (target misclassification label). This transformation path has naturally higher similarity for noise generated on samples from the same source targeting the same class, compared to noise generated on samples belonging to different pair (source,target) classes. Therefore, our intuition is that a static noise component could be present in high-dimensional space among the samples from a given class in the path towards a specific target class given a decision boundary.
\begin{figure}[!t]
\centering
\includegraphics[width=\columnwidth]{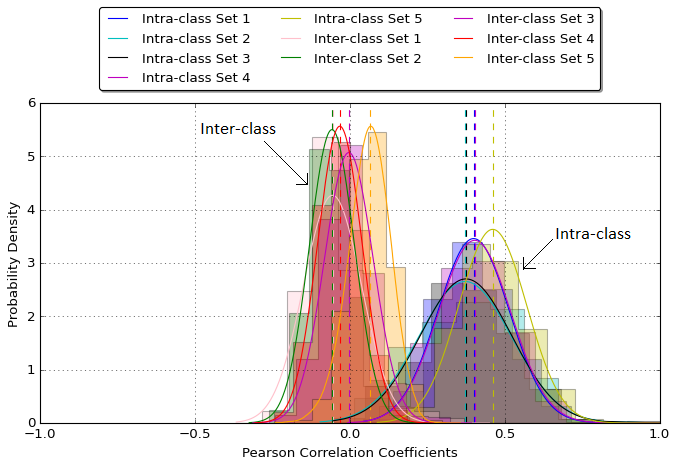}
\caption{Gaussian distribution of Pearson correlation coefficients PGD MNIST intra-class vs inter-class.}
\label{pgd_mnist_pcc}
\end{figure}

We analyze the similarities between noise distributions within a given setting (source,target), which we call intra-class similarity, comparatively with the distribution of noise similarities between two different settings (source,target), which we call inter-class similarity. 

Let $x_i$ be an input sample and $f$ a given classifier such that $f(x_i) = l$. Let be a targeted adversarial attack that generates an adversarial noise defined as: 
%\begin{align*}
\begin{equation}
    \varepsilon^{lk}_i ~. s.t. ~ f(x_i + \varepsilon^{lk}_i) = k  \neq l
\end{equation}
%\end{align*}

Let ${S}_{ij}$ be the similarity between two noises $\varepsilon^{lk}_i $ and $\varepsilon^{pq}_j$ for two samples $x_i$ and $x_j$, respectively, defined as follows:
\begin{equation}
%\begin{align*}
    \mathcal{S}_{ij} =  PCC(\varepsilon^{lk}_i , \varepsilon^{pq}_j)
\end{equation}
%\end{align*}

Where PCC is the Pearson correlation coefficient \cite{pcc}, used as a similarity metric. We call $\mathcal{S}_{ij}$ an \textbf{inter-class} similarity when $(p,q) \neq (l,k)$, and we call  $\mathcal{S}_{ij}$ an \textbf{intra-class} similarity when $(p,q) = (l,k)$.

%PCC values calculated between different noise distributions for two different scenarios: intra-class, i.e., noise distributions within the same (source,target) pair and inter-class, i.e., noise distributions within different pairs of (source,target) classes. 

%\ihsen{I need to harmonize distribution definition  --- }\\
%We define $P_{mn}$, a noise distribution, as the difference between a clean image $X_{mn}$ from source class $S_m$ and the generated adversarial example $X_{mn_{adv}}$ classified as target class $T_m$: 
%$(P_{m1}, P_{m2}, … , P_{mn}) = (X_{m1}, X_{m2}, … , X_{mn}) - (X_{m1_{adv}}, X_{m2_{adv}}, … , X_{mn_{adv}})$

% An intra-class Set corresponds to: \\
% $PCC((P_{k1}, P_{k2}, …, P_{kn}), (P_{k1}, P_{k2}, …, P_{kn}))$ \\
% An inter-class Set corresponds to: \\
% $PCC((P_{k1}, P_{k2}, …, P_{kn}), (P_{p1}, P_{p2}, …, P_{pn}))$, where ($k\neq p$)

\begin{figure}[!htp]
\centering
\includegraphics[width=\columnwidth]{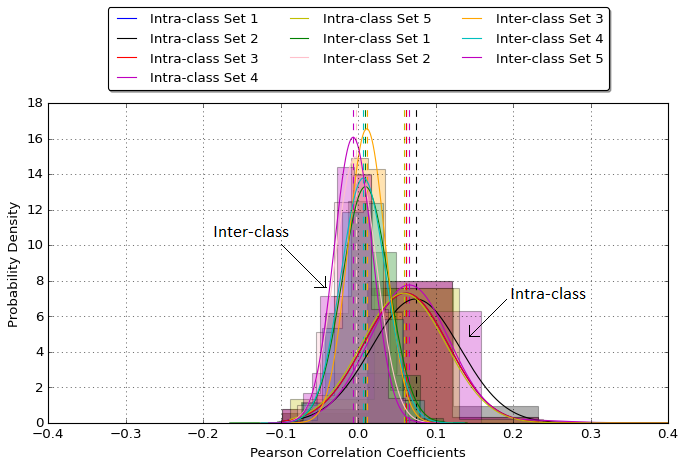}
\caption{Gaussian distribution of Pearson correlation coefficients PGD cifar-10 intra-class vs inter-class.}
\label{pgd_cifar_pcc}
\end{figure}

\begin{figure}[!htp]
\centering
\includegraphics[width=\columnwidth]{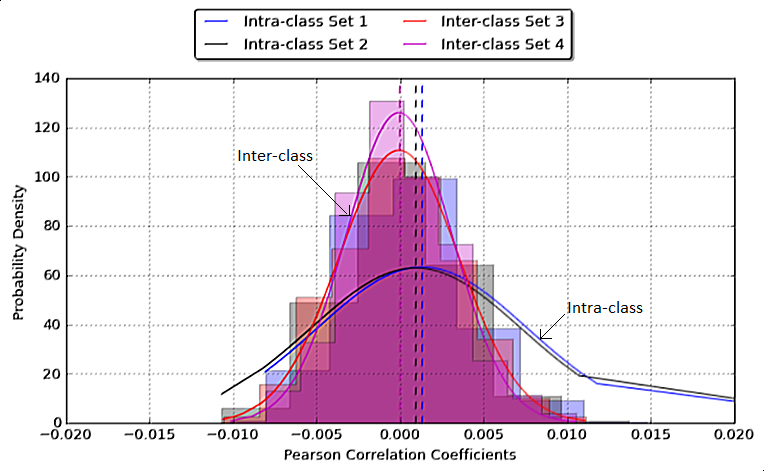}
\caption{Gaussian distribution of Pearson correlation coefficients PGD ImageNet intra-class vs inter-class.}
\label{imagenet_pcc_}
\end{figure}

In Figures \ref{pgd_mnist_pcc}, \ref{pgd_cifar_pcc} and \ref{imagenet_pcc_}, we represent the distribution of different measured similarity $\mathcal{S}_{ij}$ values for MNIST, CIFAR-10 and ImageNet datasets, respectively. We notice that the intra-class similarity distribution is clearly higher than the inter-class similarity distribution for the three datasets. This observation supports our intuition of the presence of patterns for the same (source, target) setting of adversarial noise. It, therefore, explains the mechanism by which an offline exploration allows to converge more quickly to an adversarial example. %, which  a time-aware analysis of adversarial noise generation that exploits the potential static noise patterns. 
%In Figure \ref{imagenet_pcc_}, we show the Gaussian distribution of different PCC values corresponding to patterns extracted from ImageNet images. %We notice smaller difference between distributions, this can be explained by the nature of the images (strong variety between images from the same class and higher resolution image)

%\begin{figure}
%\centering
%\includegraphics[width=0.9\columnwidth]{figures/pcc_different_number_of_iterations.png}
%\caption{Pearson correlation coefficients Gaussian distribution with an %histogram pgd mnist for different number of iterations (*).}
%\label{pcc}
%\end{figure}

%\subsection{Analysis}

\section{Discussion}
%\ihsen{put clearly the perspective of real-time, check how well can we do, what can be done --}

For the first time, this paper introduces a new problem of crafting effective AML attacks under time constraints. From a run-time perspective, existing state-of-the-art methods represent two ends of the spectrum: fully offline universal attacks and fully online input-specific attacks. To show the continuum over time of adversarial attack generation problem, we propose a new adversarial noise generation method, ROOM, that bridges the two AML approaches in the quest of fast generation of input-specific adversarial perturbations. Specifically, ROOM uses an offline-generated noise pattern, specialized for the current input, to warm up the online component of the attack. As a result, the online exploration can rapidly converge towards a successful attack by starting from this offline-generated specialized static noise, making input-specific attacks much faster.  We show that for the same time budget, ROOM substantially outperforms state-of-the-art adversarial attack methods. In fact, for the same attack success rate, ROOM converges to an adversarial example up to $5$ times and $12$ times faster on average than C\&W~\cite{C&W} and PGD~\cite{pgd} attacks, respectively. Importantly, ROOM can meet real-time constraints for a camera streaming with $25$ frames per second, generating adversarial examples at a matching rate. In fact, for a noise budget of $0.035$ and a time budget of $0.03$ seconds, ROOM achieves up to $85\%$ attack success rate, at $25$ fps throughput.  

We believe that ROOM is an important step towards producing adversarial attacks that can be deployed just-in-time. Just-in-time attacks can be a game-changer in the adversarial attacks threat model since it makes it possible for malicious actors to deploy adversarial attacks opportunistically based on the state of the environment. Additionally, these attacks are highly challenging to protect from, especially for real-time applications and Cyber-Physical Systems. Therefore, the community needs to understand the attacker's capability, which we advance using ROOM attacks. 

From another perspective, we showed that ROOM can be combined with YOPO, which is dedicated to accelerating adversarial training. This is due to ROOM's orthogonal property to gradient-based methods, and hence could be used for a more time-efficient adversarial training. 

ROOM is a general pattern of attack relying on the offline component to create a more favorable initial state for the online exploration.  We presented example implementations of the offline component that we believe can still be substantially improved.  %While OOM points out the risk of just-in-time attacks and the importance of time-aware defenses, we believe that there remains substantial room for improvement. In fact, the set of adversarial noise examples used for the offline exploration is currently randomly selected; we believe this can be enhanced with a more sophisticated approach. 
Moreover, we believe that additional optimization opportunities remain, including the use of hardware acceleration.  Finally, we would also like to explore whether the ROOM strategy results in different opportunities for defenders because of the resulting noise distribution, which is likely to be different from online only attacks.  Our future work will also explore ROOM on different learning structures and applications such as audio/NLP processing.

%\ihsen{give examples of $x \times $ less time than conventional, followed by real-time (for example common image streaming throughput (25 frame per second for example) ... etc }

% %\ihsen{Amira, could u please move this to the comments on the experimental results and highlight it in a different color to keep track? Thanks--}ok

%With this technique we still have room for improvement, for instance, because making a random selection is not a guarantee for successful AE generation (case of propose C\&W-based attack). Further enhancement in attack success rate can be attained with optimal pattern choice. 

\section{Related work}\label{sec:rw}
%\begin{itemize}
%    \item soa adversarial attacks
%    \item universal adversarial patchespatch 
%    \item Real time AA \cite{Gong2019RT} : Probably first work to formalize the problem. Using a part of the signal/data, try to predict an adversarial noise. They used Reinforcement learning for the noise generation
%    \item Real time attack against recurrent nural networks \cite{RT_rnn}: \item AdvPulse (CCS2020) \cite{advPulse}: no synchronization with the input. This is a universal acoustic patch. 
%    \item Using generative model to generate real-time adversarial attacks on automatic speech recognition \cite{aaai21}. Unpractical since the propagation time would be lower than the generative model delay added to the propagation delay of the generated adversarial noise.
%\end{itemize}

State-of-the-art adversarial attacks such as PGD \cite{pgd}, and C\&W \cite{C&W} are found to be effective in fooling DNNs. However, these attacks rely on time-consuming iterative optimization approaches, making them too slow to be launched against real-time systems.
These methods are based on the assumption of an infinite online time budget. The attacker has access to the entire data sample with a single constraint on the noise magnitude, which is not suitable for real-life situations.

Universal adversarial perturbations \cite{Universal} are based on generating an adversarial perturbation patch that works for a variety of samples. The universal adversarial perturbation is fully created offline and does not use real-time observations to improve efficiency for a target data sample. This substantially limits their efficiency.

One of the preliminary works that studied dynamic real-time adversarial attacks was \cite{Gong2019RT}, where a real-time adversarial attack for models with streaming inputs is proposed. In this scenario, an attacker can only view previous portions of the data sample and only introduce perturbations to future portions of the data sample, whereas the target model's decision will be predicated based on the entire data sample. The generated noise uses imitation learning and behavioral cloning algorithm to train real-time adversarial perturbation generator through non-real-time adversarial perturbation generator.

Li et al.~\cite{RT_rnn} use deep reinforcement learning to generate periodic adversarial perturbations to attack a recurrent neural network processing sequential data. The attack is used to generate adversarial perturbations to fool the DeepSpeech Speech Recognition system. Z. Li et al. \cite{advPulse} proposed AdvPulse: a penalty-based universal adversarial perturbation generation approach that incorporates the varying time into the optimization process to get around the constraints on speech content and time. Another work \cite{aaai21} proposed a fast audio adversarial perturbation generator (FAPG), which uses a generative model to generate adversarial perturbations for the audio input in a single forward pass. However, this method is unpractical since the propagation time would be lower than the generative model delay added to the generated adversarial noise propagation delay.

%most efforts in this direction (real-time adversarial attacks) focus on audio adversarial attacks

\section{Conclusions}
This paper contributes a new perspective of generating and analyzing adversarial attacks by introducing a real-time constraint.
We present a novel real-time online-offline attack model (ROOM) to rapidly generate adversarial attacks suitable for use in just-in-time attack settings.  ROOM leverages an offline component to support the online algorithm, allowing for rapid convergence to highly successful attacks.  %Our attack is able to achieve success similar to state of the art attacks, with an online run-time x times faster
Our results show that using an offline adversarial pattern as a starting point for the online exploration accelerated conventional adversarial attacks.
For instance, our proposed PGD-based attack achieves a $100\%$ attack success rate for a noise budget of $0.25$ under $0.0008$ seconds time constraint, whereas, for the same allocated runtime, the conventional attack is unable to generate any successful adversarial example, even for a higher noise ($0.4$) for MNIST database.

%\section*{Acknowledgment}

%\section*{References}
%\flushend 

\bibliographystyle{IEEEtran}
\bibliography{main}

\begin{thebibliography}{10}
\providecommand{\url}[1]{#1}
\csname url@samestyle\endcsname
\providecommand{\newblock}{\relax}
\providecommand{\bibinfo}[2]{#2}
\providecommand{\BIBentrySTDinterwordspacing}{\spaceskip=0pt\relax}
\providecommand{\BIBentryALTinterwordstretchfactor}{4}
\providecommand{\BIBentryALTinterwordspacing}{\spaceskip=\fontdimen2\font plus
\BIBentryALTinterwordstretchfactor\fontdimen3\font minus
  \fontdimen4\font\relax}
\providecommand{\BIBforeignlanguage}[2]{{%
\expandafter\ifx\csname l@#1\endcsname\relax
\typeout{** WARNING: IEEEtran.bst: No hyphenation pattern has been}%
\typeout{** loaded for the language `#1'. Using the pattern for}%
\typeout{** the default language instead.}%
\else
\language=\csname l@#1\endcsname
\fi
#2}}
\providecommand{\BIBdecl}{\relax}
\BIBdecl

\bibitem{simonyan2014deep}
K.~Simonyan and A.~Zisserman, ``Very deep convolutional networks for
  large-scale image recognition,'' 2014.

\bibitem{redmon2016yolo9000}
J.~Redmon and A.~Farhadi, ``Yolo9000: Better, faster, stronger,'' 2016.

\bibitem{deng2018deep}
L.~Deng and Y.~Liu, \emph{Deep learning in natural language processing}.\hskip
  1em plus 0.5em minus 0.4em\relax Springer, 2018.

\bibitem{pierson2017deep}
H.~A. Pierson and M.~S. Gashler, ``Deep learning in robotics: a review of
  recent research,'' \emph{Advanced Robotics}, vol.~31, no.~16, pp. 821--835,
  2017.

\bibitem{al2017deep}
M.~Al-Qizwini, I.~Barjasteh, H.~Al-Qassab, and H.~Radha, ``Deep learning
  algorithm for autonomous driving using googlenet,'' in \emph{2017 IEEE
  Intelligent Vehicles Symposium (IV)}.\hskip 1em plus 0.5em minus 0.4em\relax
  IEEE, 2017, pp. 89--96.

\bibitem{miotto2018deep}
R.~Miotto, F.~Wang, S.~Wang, X.~Jiang, and J.~T. Dudley, ``Deep learning for
  healthcare: review, opportunities and challenges,'' \emph{Briefings in
  bioinformatics}, vol.~19, no.~6, pp. 1236--1246, 2018.

\bibitem{b0}
\BIBentryALTinterwordspacing
F.~Tram{\`e}r, F.~Zhang, A.~Juels, M.~K. Reiter, and T.~Ristenpart, ``Stealing
  machine learning models via prediction apis,'' in \emph{25th {USENIX}
  Security Symposium ({USENIX} Security 16)}.\hskip 1em plus 0.5em minus
  0.4em\relax Austin, TX: {USENIX} Association, Aug. 2016, pp. 601--618.
  [Online]. Available:
  \url{https://www.usenix.org/conference/usenixsecurity16/technical-sessions/presentation/tramer}
\BIBentrySTDinterwordspacing

\bibitem{papernot2016transferability}
N.~Papernot, P.~McDaniel, and I.~Goodfellow, ``Transferability in machine
  learning: from phenomena to black-box attacks using adversarial samples,''
  2016.

\bibitem{fgsm}
I.~J. Goodfellow, J.~Shlens, and C.~Szegedy, ``Explaining and harnessing
  adversarial examples,'' 2014.

\bibitem{physical}
\BIBentryALTinterwordspacing
A.~Kurakin, I.~J. Goodfellow, and S.~Bengio, ``Adversarial examples in the
  physical world,'' \emph{CoRR}, vol. abs/1607.02533, 2016. [Online].
  Available: \url{http://arxiv.org/abs/1607.02533}
\BIBentrySTDinterwordspacing

\bibitem{C&W}
\BIBentryALTinterwordspacing
N.~Carlini and D.~A. Wagner, ``Towards evaluating the robustness of neural
  networks,'' \emph{CoRR}, vol. abs/1608.04644, 2016. [Online]. Available:
  \url{http://arxiv.org/abs/1608.04644}
\BIBentrySTDinterwordspacing

\bibitem{carlini_gift}
N.~Carlini, A.~Athalye, N.~Papernot, W.~Brendel, J.~Rauber, D.~Tsipras,
  I.~Goodfellow, A.~Madry, and A.~Kurakin, ``On evaluating adversarial
  robustness,'' 2019.

\bibitem{pgd}
A.~Madry, A.~Makelov, L.~Schmidt, D.~Tsipras, and A.~Vladu, ``Towards deep
  learning models resistant to adversarial attacks,'' 2017.

\bibitem{CarliniW17}
\BIBentryALTinterwordspacing
N.~Carlini and D.~A. Wagner, ``Adversarial examples are not easily detected:
  Bypassing ten detection methods,'' \emph{CoRR}, vol. abs/1705.07263, 2017.
  [Online]. Available: \url{http://arxiv.org/abs/1705.07263}
\BIBentrySTDinterwordspacing

\bibitem{distillation_SP}
N.~{Papernot}, P.~{McDaniel}, X.~{Wu}, S.~{Jha}, and A.~{Swami}, ``Distillation
  as a defense to adversarial perturbations against deep neural networks,'' in
  \emph{2016 IEEE Symposium on Security and Privacy (SP)}, May 2016, pp.
  582--597.

\bibitem{11}
T.~{Chavdarova}, P.~{Baqué}, S.~{Bouquet}, A.~{Maksai}, C.~{Jose},
  T.~{Bagautdinov}, L.~{Lettry}, P.~{Fua}, L.~{Van Gool}, and F.~{Fleuret},
  ``Wildtrack: A multi-camera hd dataset for dense unscripted pedestrian
  detection,'' in \emph{2018 IEEE/CVF Conference on Computer Vision and Pattern
  Recognition}, 2018.

\bibitem{16}
A.~{Ben Khalifa}, I.~Alouani, M.~A. Mahjoub, and A.~Rivenq, ``A novel
  multi-view pedestrian detection database for collaborative intelligent
  transportation systems,'' \emph{Future Generation Computer Systems}, vol.
  113, pp. 506--527, 2020.

\bibitem{5}
S.~{Thys}, W.~V. {Ranst}, and T.~{Goedemé}, ``Fooling automated surveillance
  cameras: Adversarial patches to attack person detection,'' in \emph{2019
  IEEE/CVF Conference on Computer Vision and Pattern Recognition Workshops
  (CVPRW)}, 2019, pp. 49--55.

\bibitem{6}
X.~Liu, H.~Yang, Z.~Liu, L.~Song, H.~Li, and Y.~Chen, ``Dpatch: An adversarial
  patch attack on object detectors,'' 2019.

\bibitem{7}
D.~Karmon, D.~Zoran, and Y.~Goldberg, ``{L}a{VAN}: Localized and visible
  adversarial noise,'' in \emph{Proceedings of the 35th International
  Conference on Machine Learning}, ser. Proceedings of Machine Learning
  Research, vol.~80.\hskip 1em plus 0.5em minus 0.4em\relax PMLR, 2018.

\bibitem{RT_rnn}
C.~R. Serrano, P.~Sylla, S.~Gao, and M.~A. Warren, ``Rta3: A real time
  adversarial attack on recurrent neural networks,'' in \emph{2020 IEEE
  Security and Privacy Workshops (SPW)}, 2020, pp. 27--33.

\bibitem{Gong2019RT}
Y.~Gong, B.~Li, C.~Poellabauer, and Y.~Shi, ``Real-time adversarial attacks,''
  \emph{ArXiv}, vol. abs/1905.13399, 2019.

\bibitem{pbform}
\BIBentryALTinterwordspacing
X.~Yuan, P.~He, Q.~Zhu, R.~R. Bhat, and X.~Li, ``Adversarial examples: Attacks
  and defenses for deep learning,'' \emph{CoRR}, vol. abs/1712.07107, 2017.
  [Online]. Available: \url{http://arxiv.org/abs/1712.07107}
\BIBentrySTDinterwordspacing

\bibitem{mnist}
\BIBentryALTinterwordspacing
Y.~LeCun and C.~Cortes, ``{MNIST} handwritten digit database,'' 2010. [Online].
  Available: \url{http://yann.lecun.com/exdb/mnist/}
\BIBentrySTDinterwordspacing

\bibitem{CIFAR}
A.~Krizhevsky, ``Learning multiple layers of features from tiny images,'' Tech.
  Rep., 2009.

\bibitem{imagenet}
J.~Deng, W.~Dong, R.~Socher, L.-J. Li, K.~Li, and L.~Fei-Fei, ``{ImageNet: A
  Large-Scale Hierarchical Image Database},'' in \emph{CVPR09}, 2009.

\bibitem{PyTorch}
A.~Paszke, S.~Gross, F.~Massa, A.~Lerer, J.~Bradbury, G.~Chanan, T.~Killeen,
  Z.~Lin, N.~Gimelshein, L.~Antiga, A.~Desmaison, A.~Köpf, E.~Yang, Z.~DeVito,
  M.~Raison, A.~Tejani, S.~Chilamkurthy, B.~Steiner, L.~Fang, J.~Bai, and
  S.~Chintala, ``Pytorch: An imperative style, high-performance deep learning
  library,'' 2019.

\bibitem{foolbox}
\BIBentryALTinterwordspacing
J.~Rauber, W.~Brendel, and M.~Bethge, ``Foolbox v0.8.0: {A} python toolbox to
  benchmark the robustness of machine learning models,'' \emph{CoRR}, vol.
  abs/1707.04131, 2017. [Online]. Available:
  \url{http://arxiv.org/abs/1707.04131}
\BIBentrySTDinterwordspacing

\bibitem{yopo}
D.~Zhang, T.~Zhang, Y.~Lu, Z.~Zhu, and B.~Dong, ``You only propagate once:
  Accelerating adversarial training via maximal principle,'' 2019.

\bibitem{pcc}
T.~Anderson, ``An introduction to multivariate statistical analysis (wiley
  series in probability and statistics),'' 2003.

\bibitem{Universal}
S.~{Moosavi-Dezfooli}, A.~{Fawzi}, O.~{Fawzi}, and P.~{Frossard}, ``Universal
  adversarial perturbations,'' in \emph{2017 IEEE Conference on Computer Vision
  and Pattern Recognition (CVPR)}, 2017, pp. 86--94.

\bibitem{advPulse}
\BIBentryALTinterwordspacing
Z.~Li, Y.~Wu, J.~Liu, Y.~Chen, and B.~Yuan, ``Advpulse: Universal,
  synchronization-free, and targeted audio adversarial attacks via subsecond
  perturbations,'' in \emph{Proceedings of the 2020 ACM SIGSAC Conference on
  Computer and Communications Security}, ser. CCS '20.\hskip 1em plus 0.5em
  minus 0.4em\relax New York, NY, USA: Association for Computing Machinery,
  2020, p. 1121–1134. [Online]. Available:
  \url{https://doi.org/10.1145/3372297.3423348}
\BIBentrySTDinterwordspacing

\bibitem{aaai21}
Y.~Xie, Z.~Li, C.~Shi, J.~Liu, Y.~Chen, and B.~Yuan, ``Enabling fast and
  universal audio adversarial attack using generative model,'' 2021.

\end{thebibliography}

\end{document}